\documentclass[useAMS,usenatbib]{mn2e}

 \usepackage{Times}

\usepackage{graphicx}
\usepackage{color}
\definecolor{red}{rgb}{1,0,0}
\definecolor{blue}{rgb}{0,0,1}

\title[High dynamic range imaging]{High dynamic range imaging by pupil single-mode filtering and remapping}
\author[G. Perrin, S. Lacour, J. Woillez and \'E. Thi\'ebaut]{G. Perrin$^{1}$\thanks{E-mail:
guy.perrin@obspm.fr}, S. Lacour$^{1}$, J. Woillez$^{2}$  and  \'E. Thi\'ebaut$^{3}$\\
$^{1}$LESIA , UMR 8109, Observatoire de Paris, 5 place Jules Janssen, 92190 Meudon, France \\
$^{2}$W.M. Keck Observatory, 65-1120 Mamalahoa Highway, Kamuela, HI 96743, USA \\
$^{3}$Centre de Recherche Astronomique de Lyon, UMR 5574, 9 avenue Charles Andr\'e, 69561 Saint-Genis-Laval, France}
\begin{document}

\date{Accepted . Received ; in original form}

\pagerange{\pageref{firstpage}--\pageref{lastpage}} \pubyear{2007}

\maketitle

\label{firstpage}

\begin{abstract}
Because of atmospheric turbulence, obtaining high angular resolution images with a high dynamic range is difficult even in the near infrared domain of wavelengths. We propose a novel technique to overcome this issue. The fundamental idea is to apply techniques developed for long baseline interferometry to the case of a single-aperture telescope. The pupil of the telescope is broken down into coherent sub-apertures each feeding a single-mode fiber. A remapping of the exit pupil allows interfering all sub-apertures non-redundantly. A diffraction-limited image with very high dynamic range is reconstructed from the fringe pattern analysis with aperture synthesis techniques, free of speckle noise. The performances of the technique are demonstrated with simulations in the visible range with an 8 meter telescope. Raw dynamic ranges of 1:$10^6$ can be obtained in only a few tens of seconds of integration time for bright objects. 

\end{abstract}


\section{Introduction}
Most astronomical sources have a small angular extension. The study of their spatial intensity distribution thus requires a high angular resolution. Among these, exoplanets require in addition very high dynamic ranges to be directly imaged near their host star. High angular resolution and high dynamic range are the key. Because of diffraction, the angular resolution of a light collector linearly increases with its diameter. Larger telescopes therefore provide larger angular resolutions. However, phase aberrations due to atmospheric turbulence prevent from reaching the diffraction limit. The adaptive optics (AO) technique, in which a deformable mirror corrects the corrugated wavefront, allows to restore the diffraction limit in spite of turbulence \citep{rousset1990}. Although an elegant and effective solution, it provides a limited sky coverage in absence of a polychromatic laser guide star and is still difficult to apply to the next generation of very large telescopes at short wavelengths. Post-processing techniques such as speckle imaging and aperture masking, allow restoring diffraction limited images from short exposure images in which turbulent fluctuations are frozen. Speckle imaging \citep{labeyrie1970} has a limited dynamic range and aperture masking \citep{haniff1987} only makes use of a small fraction of the pupil. Here we propose to combine the advantages of these two techniques and to spatially filter the beams to achieve high dynamic range diffraction limited images even with extremely large telescopes at optical and near-infrared wavelengths. We first discuss the issue of imaging through the turbulent atmosphere in the optical in Section~\ref{sec:turbulence}. We present our instrument concept in Section~\ref{sec:instrument} and discuss the anticipated performances in Section~\ref{sec:performances}. Conclusions are drawn in Section~\ref{sec:conclusions}.

\section{Imaging at optical wavelengths through atmospheric turbulence}
\label{sec:turbulence}
\begin{figure*}
\hbox{
\includegraphics[width=6.5cm , angle=-90]{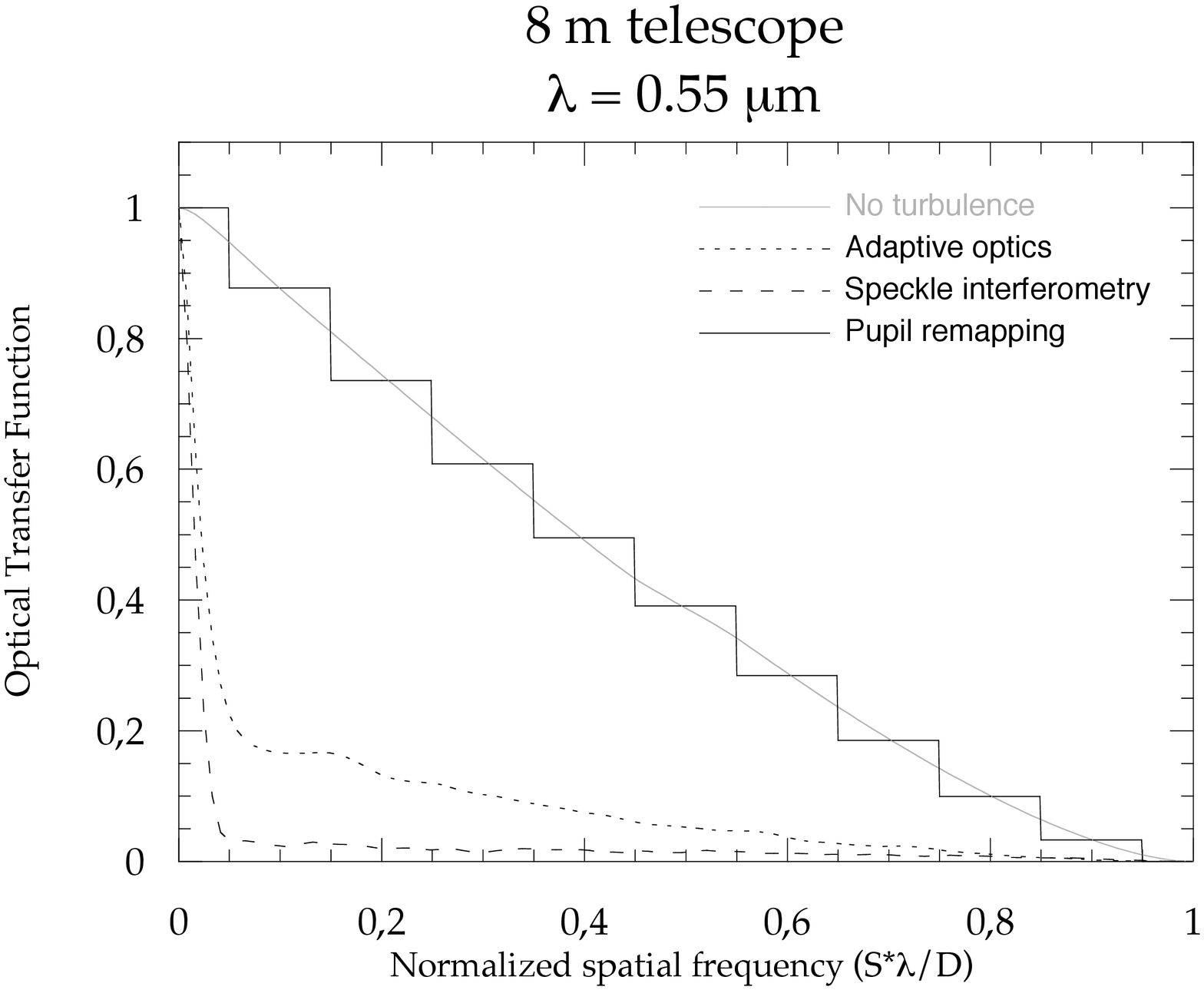}
\includegraphics[width=6.5cm , angle=-90]{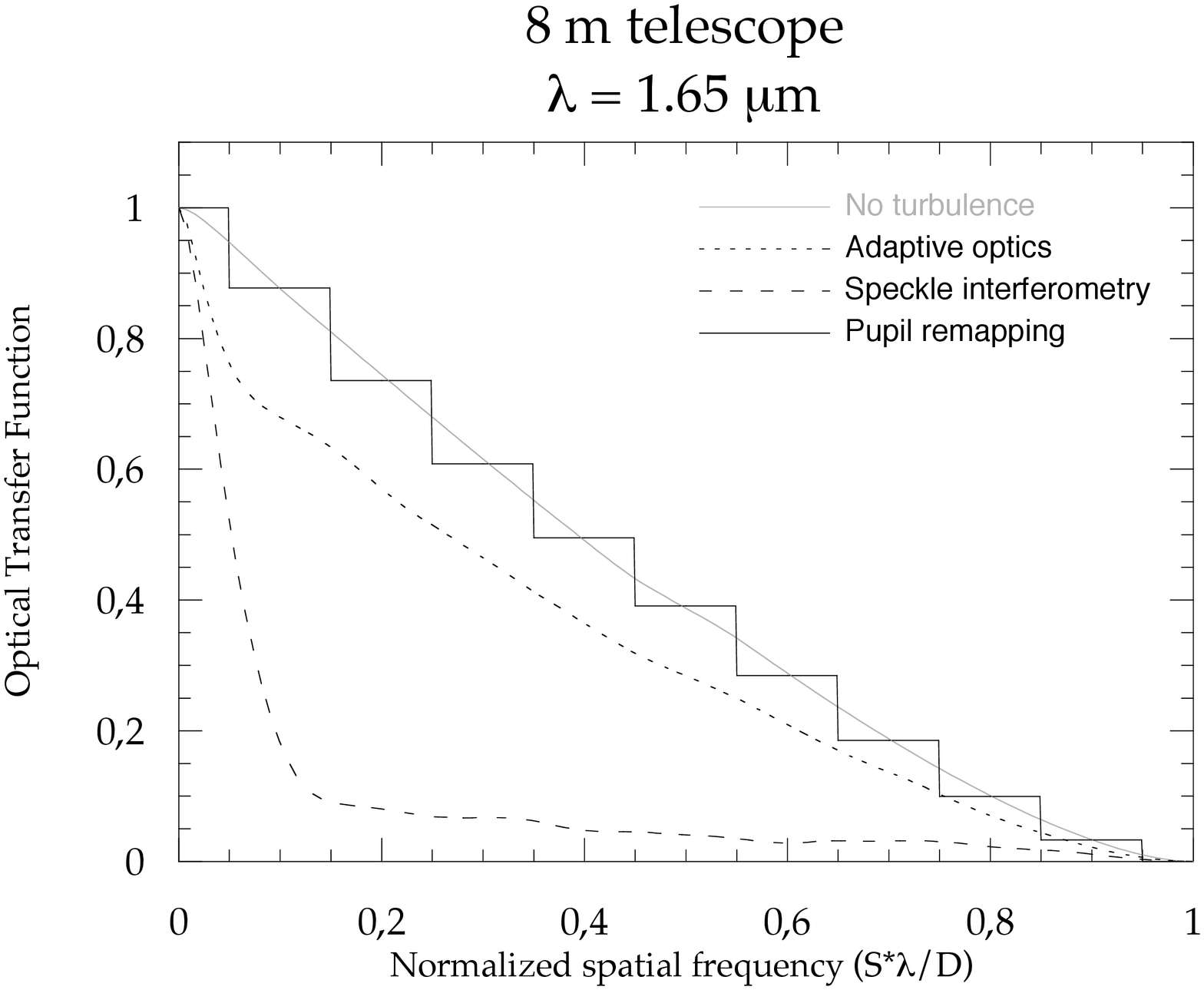}
}
\caption{8 m telescope Optical Transfer Functions. {{Grey continuous line: perfect telescope without turbulence. All three other OTFs are for excellent turbulence conditions (seeing of 0.4 arc second). Black continuous line: OTF with the system proposed in this paper. The width of the bins is proportional to the diameter of a sub-pupil. The OTF in each bin is constant and is equal to the weighted average of the filtered optical transfer function. Black dotted line: with adaptive optics correction. Black dashed line: with the speckle interferometry technique.}} Left: at a visible wavelength. Right: in the near infrared. The quality of the OTF improves with wavelength making performance at short wavelengths quite poor. With the pupil remapping system, the OTF is mechanically constant making the telescope in turbulent conditions a quasi-perfect telescope thus restoring diffraction limit and high dynamic range.}
\label{fig1}
\end{figure*}
The theory of imaging establishes that a telescope acts as a low-pass filter on an image, the optical transfer function (OTF) being a decreasing function from the zero (spatial) frequency down to the telescope cut-off frequency $D/\lambda$ with $D$ the pupil diameter and $\lambda$ the wavelength (Figure~\ref{fig1}). In presence of turbulence and for long exposures, the cut-off frequency is reduced to $r_0/\lambda$ when $r_0 < D$ hence a spatial resolution limited to $\lambda/r_0$. The Fried parameter, $r_0$, is the diameter of a coherent patch over which the variance of phase is $1~\mathrm{rad}^2$. AO partly restores the OTF up to $D/\lambda$ as long as the incoming wavefront can be analyzed on a time scale shorter than the coherence time of turbulent phase. In practice, this requirement is fulfilled at near-infrared wavelengths as the coherence time and $r_0$ increase with wavelength. Correction of turbulent phase is therefore more demanding in the visible and no large astronomical telescope has been equipped with a visible AO system yet, despite the potentially higher angular resolution. Correction of turbulent phase for future extremely large telescopes at near-infrared wavelengths will also be more difficult as the number of coherent patches will be more than an order of magnitudes larger. With the speckle technique, the OTF is also restored up to the same limit $D/\lambda$. In both cases however, the higher the spatial frequency, the larger the energy loss. This is due to the larger decorrelation of turbulent phases between the more distant parts of the pupil. 

OTFs obtained in different conditions with an 8~m telescope have been simulated and are presented in Figure~\ref{fig1}. Simulations were made in the visible ($\lambda=0.55~\mu\mathrm{m}$) and in the near infrared ($\lambda=1.65~\mu\mathrm{m}$). For turbulent conditions, an excellent seeing of 0.4~arcsecond has been chosen. The OTF of a perfect telescope (no turbulence and no aberrations) is compared to the OTFs obtained with adaptive optics, with the speckle technique and with the system proposed in this paper (the width of the bins is the width of the sub-pupils). The performance of adaptive optics and speckle interferometry are highly turbulence-dependent and therefore highly time and wavelength-dependent. In practice, turbulent conditions can change over short time scales making the calibration of the OTF difficult. The situation improves with wavelength making performance at short wavelengths quite poor. With the pupil remapping system, the OTF is mechanically constant making the telescope in turbulent conditions a quasi-perfect telescope thus restoring diffraction limit and high dynamic range (see Section~\ref{sec:instrument}).

The losses of coherence are responsible for speckle noise in the image that limits the dynamic range with both speckle imaging (to about 1:100) and AO making it difficult to reach 1:$10^5$ dynamic ranges required for exoplanetary system imaging \citep{chelli2005,cavarroc2006}. They are also responsible of a smaller contrast in the image on smaller spatial scales. In the classical imaging approach, the pupil is massively redundant as a same vector can be drawn between several pairs of points in the pupil (a given spatial frequency is measured different times in the pupil). In the OTF, a frequency component is the sum of vectors of complex amplitude produced by different pairs whose phases are the phase differences between two points of the incoming wavefront. The degradation of the OTF is therefore induced by these incoherent additions in a redundant pupil. The technique of non-redundant aperture masking overcomes this issue: a mask is placed in a pupil plane to select a set of sub-pupils organized as a non-redundant array. In the instantaneous OTF, each discrete spatial frequency is therefore generated by a single sub-aperture pair and there is no coherence loss when the sub-pupils are point-like or far smaller than $r_0$. As a consequence, the OTF is a set of peaks of unit amplitude beside the zero frequency peak. An image can then be reconstructed through the closure phase technique. This technique has been demonstrated on the sky on several telescopes and spectacular astronomical results have been obtained such as the imaging and discovery of the pinwheel nebula \citep{tuthill1999}.

\begin{figure*}
\includegraphics[width=\textwidth,angle=0]{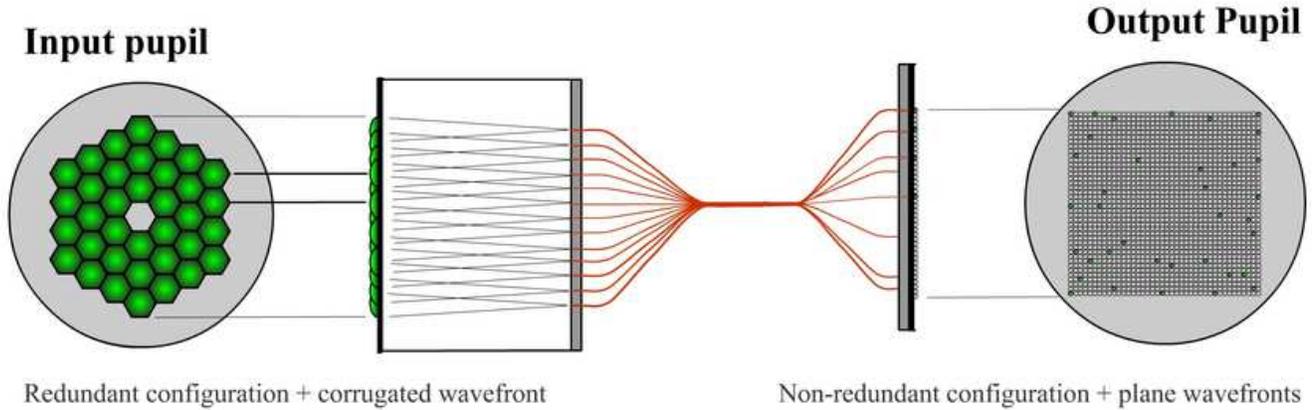}
\caption{Instrument set-up. The input pupil is mapped with a lenslet array. Each lenslet focuses a sub-pupil beam into a single-mode fiber. In the output pupil, the fibers are placed at the focus of another lenslet array. Fibers are distributed in a non-redundant array. Each pair of sub-pupils in the input pupil yields a single spatial frequency component in the output pupil. Spatial frequency components with different turbulent phases are therefore not mixed thus preserving optimum imaging performances. }
\label{fig2}
\end{figure*}

\section{Pupil remapping and spatial filtering with single-mode fibers}
\label{sec:instrument}
The shortcoming of non-redundant aperture masking is that only a small amount of the pupil can be used thus leading to a very poor sensitivity. The diameter of the sub-pupils therefore needs to be increased as much as possible to about $r_0$. This however reduces the average coherent energy per frequency bin to about 36\% and the OTF is fluctuating and difficult to calibrate. This yields a degraded dynamic range in the reconstructed image. 
This is a well-known issue in long baseline interferometry. It has been solved with single-mode fibers. Their fundamental property is that the spatial distribution of the lightwave phase is by nature flat across a waveguide section and the transported lightwave is therefore by definition perfectly coherent. The FLUOR instrument \citep{foresto1998}, based on this technology, has demonstrated that the coherent energy in long-baseline interferometry is fully restored leading to more accurate visibility measurements and to potentially higher dynamic range reconstructed images. The key is that the phase corrugations over each pupil of the interferometer are traded against intensity fluctuations which can be monitored and calibrated at the pace of turbulence. The same principle can be applied to aperture masking. The lightwave from each sub-pupil feeds a single-mode fiber. The net effect is that the frequency peaks in the OTF of a non redundant array are automatically of unit amplitude whatever the size of the sub-apertures. The use of fibers in non-redundant aperture masking has been proposed \citep{chang1998}. 
 
We propose to further take advantage of the use of fibers to rearrange the distribution of sub-apertures in the exit pupil of the instrument (Figure~\ref{fig2}). The input pupil is mapped with a lenslet array. Each lenslet focuses a sub-pupil beam into a single-mode fiber. The focal length of lenslets is chosen for optimum flux injection, the maximum coupling efficiency is 78\% for a step-index fiber. The full entrance pupil of the telescope can now be used and turned into a non-redundant array in the exit pupil of the instrument. In the output pupil, the fibers are placed at the focus of another lenslet array and recast into a non-redundant array. Each pair of sub-pupils in the entrance pupil of the telescope is associated to a single sub-pupil vector in the exit pupil. The different phases of redundant pairs are therefore not mixed by the interferometric combination and the full coherent energy is preserved. The OTF modulus in the exit pupil is equal to 1 for each exit spatial frequency or, equivalently, the OTF in the entrance pupil is maximum for each bin (Figure~\ref{fig1}). As far as the source is concerned, the measurement of the modulus of each spatial frequency component is unbiased by speckle noise and limited by photon and detector noise only. The rotation shearing interferometer \citep{mariotti1992} and the coding of non-redundant sub-arrays with differential wavefront tips and tilts \citep{monnier2004} are an alternative to achieve non-redundant pupil recombinations. However, both setups do not include spatial filtering and the achievable dynamic range is therefore limited. 

\section{Performances of the technique}
\label{sec:performances}
Although modal filtering by fibers imposes a flat wavefront in each sub-pupil, the average differential phases between sub-apertures are preserved and the exit wavefront is a step function. The closure phase technique allows eliminating these phase errors as the sum of phases over a triangle of sub-apertures is not affected by any sub-aperture phase error. It provides a fraction of $1-2/N$ ($N$ is the number of sub-apertures) of the phase information.  In our case however, the high redundancy of the entrance pupil is useful. Closure phases measured in the exit pupil yield the above fraction of phase information. Under high entrance redundancy, the number of exit closure phases is larger than the number of independent phases therefore providing the full phase information. This can be understood with a system as simple as {{$N=6$}} hexagonal-shaped sub-pupils arranged on a hexagon. The number of pairs is {{ $N_\mathrm{pairs}=N(N-1)/2=15$ }} and the number of independent triangles is { { $N_\mathrm{triangles}=(N-1)(N-2)/2=10$ }}. However, most pairs form a same vector and the number of different vectors is equal to 9. In fact, the number of measured closure phases is larger than the amount of available object phase information. This effect increases with the number of sub-apertures in such a compact array. Thanks to the redundancy, the flux coupled in each fiber can also be measured \citep{lacour2006}. Eventually, our system allows measuring all complex spatial frequency components of the object as sampled by the array of sub-pupils. The accuracy on moduli and phases can be directly derived from the achievement of single-mode long baseline interferometers: a fraction of a percent for moduli and better than a degree for phases. Better performances can be expected with a single-telescope experiment. Using conservative $\delta V = 1\%$   and $\delta \phi = 1^{o}$, an estimate of the dynamic range is obtained from the formula \citep{baldwin2002}:

\begin{equation}
DR=\sqrt{\frac{N_{pairs}}{\left( \delta V / V \right)^2+\delta\phi^{2}}}.
\end{equation}
 
\noindent For an unresolved object ($V=1$), the dynamic range is $DR\approx50\times\sqrt{N_{pairs}}$. Assuming a 132 sub-pupil array, a dynamic range of 1:4600 is achieved per snapshot. Since speckle noise has been eliminated, different snapshots are uncorrelated and the dynamic range for $S$ snapshots is equal to {{1:$4.6\times10^{3}\times \sqrt{S}$ }}. With $10^4$ snapshots, a dynamic range of 1:$4.6\times 10^{5}$  is obtained in the reconstructed image. The relevance of this formula is illustrated by Figure~3. The observation of a multi-component object under good seeing conditions (0.6 arc seconds) with a 132-element pupil remapping system at the focus of an 8 meter telescope at 630 nm wavelength (bandwidth = 60 nm) has been simulated. No adaptive optics correction is applied and only 6\% of the flux is coupled in the fibers in average. Performances in injected flux would at least double with an upstream near-infrared optimized adaptive optics correction. The object is a star surrounded by a disk whose integrated flux is a hundredth of that of the star. A decreasing exponential law has been chosen for the disk flux that is only a 1000$^\mathrm{th}$ of the star brightness at a distance equal to three times the diffraction limit. Two planets orbit the star and are dimmer by factors of $10^3$ (Planet A) and $10^4$ (Planet B) with respect to the central object. The closest planet { { (Planet A) }} is located 6 diffraction limits away from the star. The star image is saturated in the displayed images and the diffraction limit can be assessed on an unresolved planet image. The magnitude of the star decreases from the left to the right by increments of 5 in magnitude (a factor of 100 in brightness, stars from the left to the right have visual magnitudes of 10, 5 and 0, respectively). Each image is the sum of 10000 snapshots of 4~ms amounting to a total exposure time of 40 s. { {The frame rate has been chosen so that the piston is frozen during an acquisition to prevent the fringe pattern to be blurred and to keep the fringe contrast larger than 99\% (fringe contrast accuracy better than 1\%). Using the parameters of the simulations, the maximum allowed frame time is 12~ms. Should the frame rate be smaller than 100~Hz then the final dynamic range could be smaller than those achieved here. Techniques to measure the contrast loss due to piston could possibly be developed to relax this constraint and allow larger individual exposure times.}} Images are recorded in the photon noise limited regime with a photon counting camera. Images are reconstructed from the interferometric data with the algorithm of \citet{lacour2006}. The dynamic ranges directly measured in the reconstructed images are of 1:$10^4$, 1:$10^5$ and 1:$10^6$ as predicted (thanks to twice more accurate visibilities than in the above calculation). The faintest component is clearly detected in the V=5 and 0 images. \citet{lacour2006} demonstrate that speckle noise is eliminated and detection is photon and detector noise limited as anticipated when using single-mode fibers and measuring the differential phases between sub-pupils. This is where the gain in dynamic range comes from. The dynamic range can therefore still be increased by accumulating more images to reduce the impact of the noise.  
\begin{figure*}
\label{fig3}
\vbox
{
\begin{tabular}{lllll}
\hspace{0cm}V=10 & \hspace{4.4cm}V=5 & \hspace{4.4cm} & V=0 &  \\
\end{tabular}
\begin{tabular}{lllll}
& & \hspace{9.2cm} Planet B $\downarrow$ &  & \hspace{-0.1cm} Planet A $\downarrow$ \\
\end{tabular}
}
{
\includegraphics[width=17.5cm , angle=0]{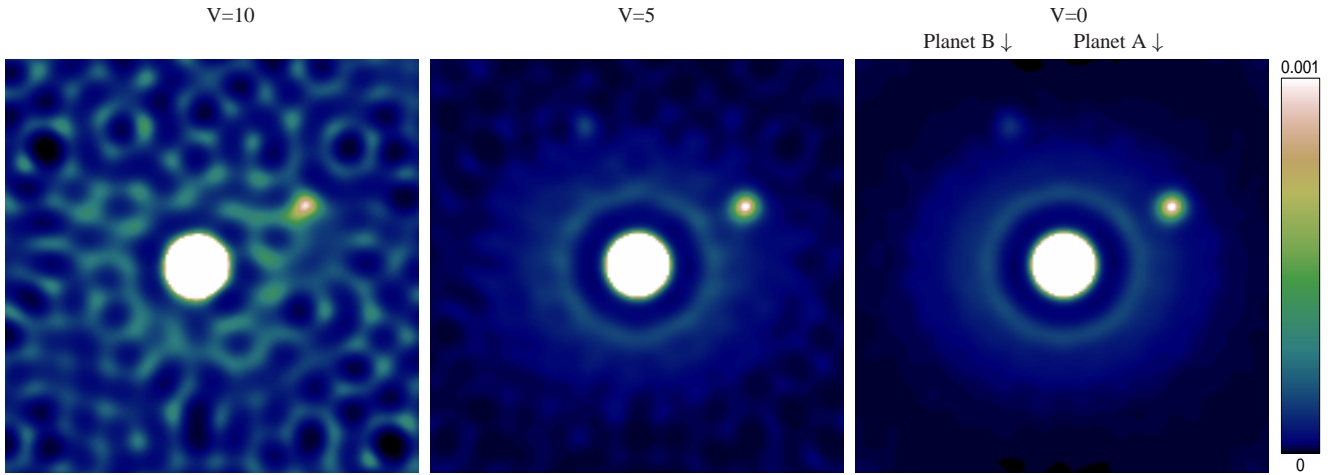}
\caption{Simulated observation and image reconstruction. Simulations of the observation of a multi-component object under good seeing conditions (0.6 arc seconds) with a 132-element pupil remapping system at the focus of an 8 meter telescope at 630 nm wavelength (bandwidth = 60 nm). No adaptive optics correction is applied. The object is a star surrounded by a disk whose integrated flux is a hundredth of that of the star. A decreasing exponential law has been chosen for the disk flux. Two planets orbit the star and are dimmer by factors of $10^3$ (Planet~A) and $10^4$ (Planet~B) with respect to the central object. The closest planet is located 6 diffraction limits away from the star. The star image is saturated and the diffraction limit can be assessed on an unresolved planet image. The magnitude of the star is respectively V = 0, 5, 10 from right to left. Each image is the sum of 10000 snapshots of 4 ms amounting to a total exposure time of 40 s. Images are recorded in the photon noise limited regime with a photon counting camera. The dynamic ranges directly measured in the images are of 1:$10^4$, 1:$10^5$ and 1:$10^6$ as predicted. The faintest component is clearly detected in the V=5 and 0 images. The speckle noise has been eliminated by the instrument and detection is photon noise limited.}
}
\end{figure*}

{ {These performances do not take into account systematics that could impair the actual performance of the instrument. As long as the bandwidth is small enough (i.e. the coherence length of the radiation is larger than the rms piston) the instrument is immune to bandwidth (chromatic) effects, piston and higher order turbulence effects. Problems may arise from imperfections in the realization of the optics and in the alignment of the instrument such as polarization and dispersion whose effects will reduce the fringe contrast and alter the fringe phase. However such static effects can be mitigated by calibrating the data with data acquired on a calibrator star unresolved by the instrument. First experiments will allow to study and discover those systematics and measure their impact on the technique.}}

Since the image is reconstructed from the measurement of spatial frequency components, the noise is white in the final image. This is a drawback with respect to direct imaging techniques for which the source photon noise is essentially concentrated in the central star image and can be rejected thanks to a coronograph. A nulling technique can however be used by applying a phase pattern to the fiber network to cancel light wherever needed. The coronograph is however of no use for multiple bright sources and for extended objects whereas our technique can be applied to any astronomical source. The advantage of our technique is the perfection of the OTF hence the absence of speckle noise. A better detectivity, or equivalently a higher dynamic range, could be achieved in different ways. The point spread function can be accurately modeled and subtracted from the reconstructed images and advanced faint object algorithms could be used to detect planets. Alternatively, the visibility data can be directly fitted with a model of the object without reconstructing an image to search for faint planets around stars to avoid reconstruction noise (the spatial frequency plane coverage is not perfect).

The high dynamic range can be as well obtained on more extended objects. {{ The instantaneous field of view is primarily limited by two effects: the lobe of each single-mode fiber matched to $\lambda/d$ where $d$ is the diameter of a sub-pupil ; and the interferometric field of view. The latter comes from the non-homothetic relation between the entrance and exit pupils of the instrument. In usual non-Fizeau combination schemes (i.e. for which the exit and entrance pupils are not homothetic), the exit pupil is densified with respect to the entrance pupil to reduce the number of peaks (or fringes) in the fringe pattern to concentrate the coherent flux in the central peak of width $\lambda/D$ with $D$ the size of the entrance pupil. The most extreme case of such a set-up is the hyper-telescope of \citet{labeyrie1996}. In this case, the interferometric field of view is a fraction $\gamma$  of $\lambda/d$ with $\gamma$ the densification factor. In the present instrument, the pupil is not densified but diluted by a large amount to reach a non-redundant configuration. The exit diffraction pattern is composed of peaks whose characteristic size is $\lambda/B_{\mathrm{max}}$ with $B_{\mathrm{max}}$ the longest baseline in the exit pupil. Those peaks are spread over a single sub-aperture diffraction pattern of width $\lambda/d$~\footnote{This can be easily demonstrated e.g. in the case of a compact square array of square sub-apertures. When applying a magnification factor to the grid spacing of the array of sub-aperture centers, peaks appear for factors larger than 1 whereas there is a single peak for a magnification factor of 1 (i.e. no magnification). This can be generalized to a non-regular array of sub-apertures as is the case for a 2-D non-redundant array but is beyond the scope of this paper. An example of such diffraction pattern can be found in Figure~3 of \citet{tuthill2000}.}. The interferometric field of view then depends upon spectral bandwidth. As a matter of fact, the wave carrier of the fringe pattern has a width at least equal to that of the diffraction pattern of a sub-pupil. For an off-axis object (off-axis distance $\alpha$) to contribute to the fringe pattern of the on-axis source, it is necessary that the coherence length of the radiation in baseline units (i.e. divided by the maximum baseline $D$) be as large as $\alpha$. Hence the classical formula for the field of view of a co-axial beamcombiner:
\begin{equation}
\alpha \leqslant \frac{\lambda^2}{D\Delta \lambda}.
\end{equation}
The field of view is maximum (i.e. the interferometric field of view matches the fiber lobe) when:
\begin{equation}
\Delta \lambda = \left (  \frac{d}{D} \right ) \times \lambda.
\end{equation}
For the simulation presented in this paper, this holds if the spectral bandwidth is roughly equal to 60~nm hence our choice. The field of view can therefore be adjusted according to the characteristics of the astrophysical targets. 
}}
If needed, the field of view can still be increased by a 2-D mapping of the sky { {if a field of view larger than the fiber lobe is necessary.}}

The use of an upstream near-infrared optimized AO is very advantageous as it allows to increase the size of each sub-pupil and therefore the sensitivity. The single-mode aperture masking imager can therefore be used on the largest telescopes at visible wavelengths behind a near-infrared optimized AO or on future extremely large telescopes either at visible or infrared wavelengths and will be able to provide very high angular resolution and dynamic range images. Last but not least, since the wavefront after filtering with fibers is a step function, the imager could also actively control wavefront errors, the actuators being fiber stretchers. In this active mode, the instrument could be a simpler AO version for future extremely large telescopes.

\section{Conclusion}
\label{sec:conclusions}

We have proposed a novel technique to obtain high angular resolution and high dynamic range images through a turbulent atmosphere whatever the wavelength, be it in the visible or in the infrared domains. The telescope pupil is broken into sub-pupils each injected into single-mode fibers to perfectly spatially filter the beams corrugated by atmospheric turbulence. A non-redundant output pupil is formed to avoid mixing phases from different parts of the input pupil. We have shown that the optical transfer function of a telescope could be restored quasi perfectly with this technique and that speckle noise is eliminated. Raw dynamic ranges of 1:$10^6$ can be obtained on bright sources in a few tens of seconds of integration time. This makes the technique very promising for the direct imaging of exoplanets around nearby stars but also for the use of current ground based telescopes and future extremely large telescopes at very high angular resolution even at short wavelengths.

\section*{Acknowledgments}

This research has made used of the YAO (http://www.maumae.net/yao/) adaptive optics simulation software developped by Franois Rigaut.

\bsp

\label{lastpage}

\end{document}